\documentclass[9pt,twocolumn,twoside]{optica}
\setboolean{shortarticle}{true}
\setboolean{minireview}{false}

\title{Off-Axis Spiral Phase Mirrors for Generating High Intensity Optical Vortices}

\author[1,*]{Andrew Longman}
\author[2,3]{Carlos Salgado}
\author[2,3]{Ghassan Zeraouli}
\author[2]{Jon I. Api\~{n}aniz}
\author[2]{Jose Antonio P\'{e}rez-Hern\'{a}ndez}
\author[1]{M. Khairy Eltahlawy}
\author[2,3]{Luca Volpe}
\author[1]{Robert Fedosejevs}

\affil[1]{Department of Electrical and Computer Engineering, University of Alberta, Edmonton, Canada, T6G1R1}
\affil[2]{CLPU, Centro de L\'{a}seres Pulsados, Building M5, Science Park, Calle Adaja 8, 37185 Villamayor, Salamanca, Spain}
\affil[3]{Universidad de Salamanca, Patio de Escuelas 1, 37008 Salamanca, Spain}

\affil[*]{Corresponding author: longman@ualberta.ca}

\begin{abstract}
In this work, we present a novel and practical method for generating optical vortices in high-power laser systems. Off-axis spiral phase mirrors are used at oblique angles of incidence in the beam path after amplification and compression allowing for the generation of high-power optical vortices in almost any laser system. An off-axis configuration is possible via modification of the azimuthal gradient of the spiral phase helix and is demonstrated with a simple model using a discrete spiral staircase. This work presents the design, fabrication, and implementation of off-axis spiral phase mirrors in both low and high-power laser systems.
\end{abstract}

\setboolean{displaycopyright}{true}

\begin{document}

\maketitle

\section{Introduction}

The generation and application of laser beams carrying orbital angular momentum (OAM) has become one of the forefront research areas of laser engineering today. Laser-matter interactions with beams carrying OAM at relativistic intensities is an area of recent study and has been drawing attention from the laser-plasma community. While there has been significant theoretical interest \cite{PhysRevLett.112.215001,doi:10.1063/1.5044617,Amplificationraman}, there have been few experimental studies \cite{Brabetz,plasmaholograms,PhysRevLett.118.033902} due to difficulties associated with generating OAM beams in high-power laser systems.   

Classically, an electromagnetic wave carrying OAM propagates with a helical wavefront contrary to the planar wavefronts of most high-powered lasers today. The pitch of the helical wavefront $Q$, indicates the quantity of OAM in the beam and is typically an integer multiple of the beam wavelength $Q=L\lambda$ where $L$ is the topological charge.  Imprinting a helical wavefront into the beam gives an additional parameter for controlling both existing and new regimes of laser-matter interactions. One of the more intriguing properties of OAM beams is the unbounded angular momentum density, that is, they can carry any amount of OAM per photon \cite{PhysRevA.45.8185}. This contrasts with circularly polarized beams which are limited to contain no more than one unit of spin angular momentum $\hbar$ per photon. Coupling additional photon angular momentum to a plasma allows for new phenomena such as the generation of relativistic azimuthal electron currents \cite{PhysRevLett.112.215001}, strong axial magnetic fields \cite{PhysRevLett.105.035001}, and twisted betatron radiation \cite{PhysRevLett.118.094801}, to name just a few.
 
Spiral phase plates (SPP’s) \cite{BEIJERSBERGEN1994321,Sueda:04} are a simple and economical method for generating OAM beams. Their ability to generate high purity, low-power OAM modes is limited only by the wavefront quality, pulse duration, and near-field spatial homogeneity. High-power, short-pulse laser systems face problems with SPP’s as the transmitted wavefront is altered temporally and spatially due to non-linear effects such as group velocity dispersion, self-phase modulation and self-focusing. Alternatively, OAM beams may be generated through q-plates and cylindrical mode converters but require specialized polarization and input laser modes not generally available in high-power laser systems \cite{PhysRevA.45.8185}. Spatial light modulators are a diffractive optical element providing a flexible method for generating OAM beams as they allow for \textit{in-situ} adjustment of the diffraction pattern enabling fine tuning of the OAM mode in real time. While suitable for low-power laser systems, their damage threshold is too low for the fluence of high-power lasers. Spiral phase mirrors (SPM’s) were introduced in normal incidence configurations to overcome some of the nonlinear issues associated with SPP's \cite{Campbell:12}, but are restricted for use in high-powered lasers as they can retro-reflect damaging amounts of energy into the laser amplifier and compressor.  

There have been a few successful experiments published in which an OAM mode has been demonstrated at high-power. In one of these, a spiral phase plate was inserted in the low-power front end of the laser system yielding an asymmetric OAM mode after amplification \cite{Brabetz}. Another used various methods to generate high-power OAM beams including plasma holograms \cite{plasmaholograms}, a SPP, and a diffractive fork grating after amplification \cite{PhysRevLett.118.033902}. While these methods were successful to generate high-power OAM beams, their implementation is not straightforward for many high-power laser systems, particularly those with large diameter, short pulse, or high energy beams. To experimentally realize OAM beams at high-power, an ideal device would be inserted into a beam line after amplification and compression where the ejected super-Gaussian or flat-top beam is then mode converted into an OAM beam with maximal conversion efficiency, and with minimal spatio-temporal beam aberrations. This device would require a high damage threshold and the flexibility to be manufactured for large diameter beams with minimal cost. 

In this paper we introduce the concept of an off-axis spiral phase mirror (OASPM) that enables conversion of fundamental laser modes to OAM modes in almost any high-power laser system. The OASPM is successfully demonstrated in both low and high-power laser systems with high conversion efficiency and high symmetry beams. The experimental results are compared to theoretical calculations in the paraxial limit.

\section{Off-axis spiral phase mirrors}

To transform a spiral phase mirror (SPM) to the off-axis case, we consider a stepped spiral phase mirror as shown in Figs. 1(a) and 1(c) \cite{Sueda:04,Campbell:12}. A stepped SPM uses a helical spiral staircase imprinted on the mirror surface that transfers OAM to the beam through reflection. Discretization of the spiral into a staircase-like surface simplifies manufacturing while having a minimal impact on the OAM beam quality \cite{Longman:17,Sueda:04}. Fig 1(c). shows a stepped SPM with 16 steps of equal angular spacing for the normal incidence case, a red dotted circle is used to illustrate the laser spot imprint on the mirror. The azimuthal angle of each step sector measured from the horizontal plane can be calculated as $\Delta\phi = 2\pi/N$, given the total number of steps $N$.

\begin{figure} \label{fig:1}
	\includegraphics{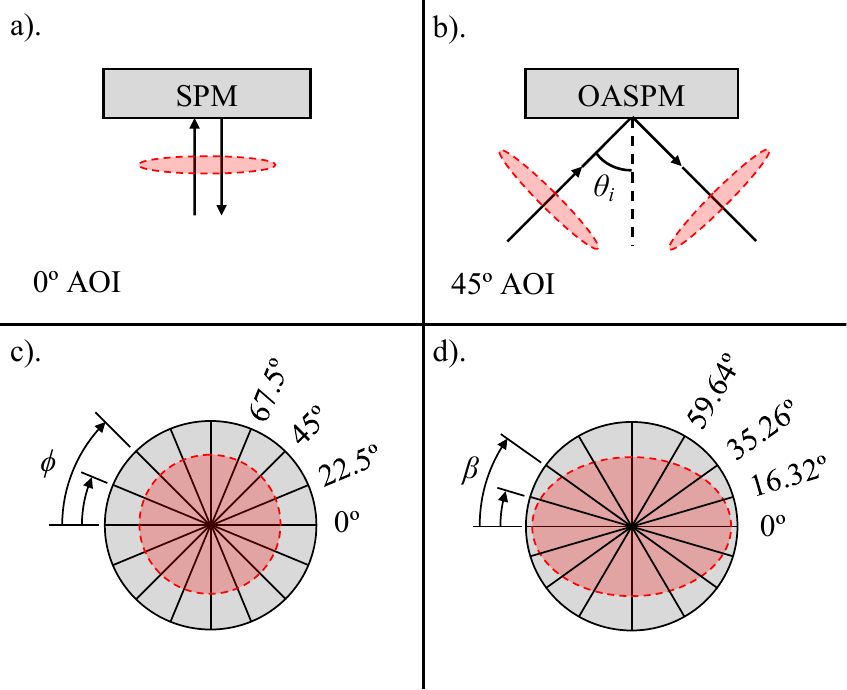}
	\caption{a) The retro-reflecting configuration of a normal incidence (SPM); b) The oblique angle of incidence to an (OASPM), shown here at $45^\circ$; c) Front view of a stepped SPM indicating the step angle $\phi$ from horizontal and circular laser beam outline in red; d) Front view of a stepped OASPM indicating the step angle $\beta$ relative to the horizontal plane and the elliptical laser beam outline in red.}
\end{figure}

If the laser is incident in the horizontal plane at an angle $\theta_i$, the laser spot imprint is transformed from a circle to an ellipse as shown in Figs. 1(b) and 1(d). Using a stepped SPM enables easy visualization of the requirement that each step sector within the laser spot ellipse must have an equal area. If we define each step sector angle from the horizontal axis as $\beta$, then it is straightforward to derive the off-axis step sector angle as a function of incidence angle and normal step sector angle,  
\begin{equation} \label{eq:12}
\beta = \arctan(\cos(\theta_i)\tan(\phi))
\end{equation}This formula gives the necessary transformation to convert the step sector angle relative to the horizontal plane of a normal incidence stepped SPM, to the off-axis case. The differential limit of this equation can be used for generating continuous (infinite step number) OASPM's. The standard incident angle case of $45^\circ$ is computed and given in Fig 1(d) for a 16 step OASPM. Experimentally implementing these variable step sector sizes is straightforward using standard material deposition techniques such as sputtering or electron beam evaporation and the mask method outlined in Sueda et al. \cite{Sueda:04}.

A second variable to consider when designing an OASPM is the staircase step height $h$, and the $2\pi$ phase discontinuity height $H$ shown in Fig. 2. For a normal incidence, continuous SPM, $H$ is equal to half of the desired OAM helical pitch $H=\lambda L/2$. At oblique angles of incidence, we must consider the additional transverse distance the beam will travel to give the net phase shift given by the Bragg reflection criterion. If we consider the diagram in Fig. 2 which uses 2D cylindrical co-ordinates, we can visualize the geometries involved. Considering the case of a continuous OASPM indicated by the dashed green line in Fig. 2, we geometrically derive the relationship between step height $H$ and the incident angle as a function of laser wavelength $\lambda L = 2H\cos(\theta_i)$. Further generalization of this equation to include $N$ discrete steps of height $h$ gives,
\begin{equation} \label{eq:14}
h = \frac{\lambda L}{2N\cos(\theta_i)}
\end{equation}
\begin{figure} \label{figure:2}
	\includegraphics{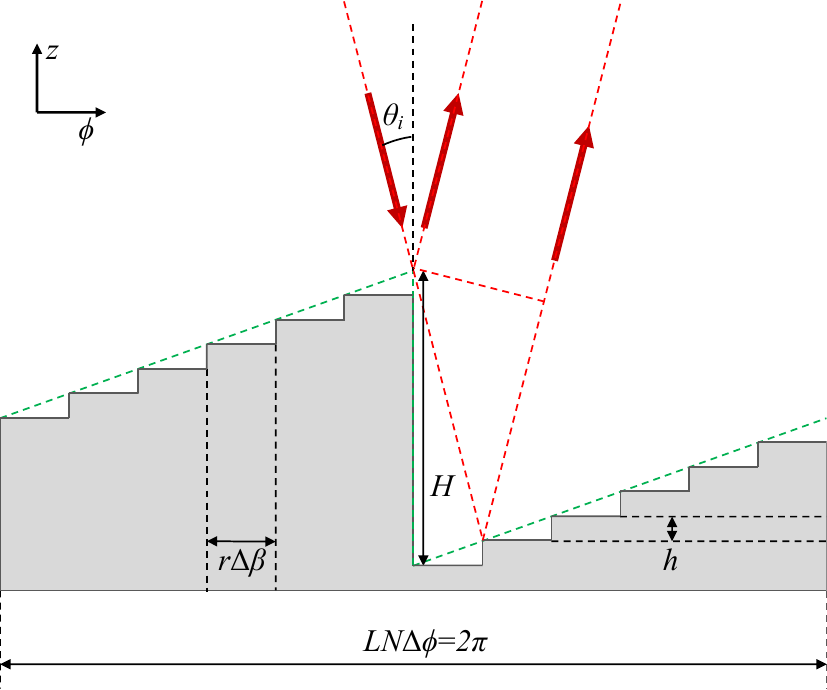}
	\caption{Illustration of an $L=1$, $N=12$ stepped OASPM in $\phi-z$ coordinates. The green dotted line represents a continuous OASPM ($N\rightarrow\infty$), and the red lines indicate light rays at a given incidence angle $\theta_i$.}
\end{figure}

Though using a finite number of steps fails to create a total phase step of $2\pi$, it has been shown that a stepped SPM with at least 16 steps per $L$ leads to a minimal reduction in the mode conversion quality, and increasing the step number beyond 32 steps per $L$ has a negligible return on beam quality \cite{Longman:17}.

From Fig. 2, we can identify a region where the incident beam is shadowed from interacting with the step adjacent to the $2\pi$ phase step. This can be mitigated by manufacturing the OASPM such that the $2\pi$ phase step is parallel to the plane of incidence. The maximum shadowing expected to occur will then be from the individual steps which are on the order of $\lambda/N$ in height. At $\theta_i = 45^\circ$, we expect the shadowed area to be equal to the step height and is therefore negligible compared to the wavelength of light. An additional limitation of the stepped OASPM is diffraction from the step edges depending on the manufacturing technique and the step edge width. Through fabrication of the OASPM via electron beam evaporation, we found the step edge width to be of the order of 100$\mu m$ primarily as a result of mask position uncertainty as masks are interchanged. The corresponding area of the step edges relative to the total surface area of a large diameter beam $(>100mm)$ is around 1$\%$ and assumed to have a negligible effect on the focal spot. Much sharper edges could be produced using microelectronic lithographic techniques and would be required for microscale OASPM's. Sharp step edges may however generate high local electric fields that can either heat or damage the OAPSM surface with sufficient laser fluence. This could be mitigated through the use of high reflectivity dielectric coatings or by grading the step edges over several wavelengths reducing any sharp electric field spikes while still only contributing to very small diffraction losses at the steps. Prototype OASPM's using unprotected gold spiral staircases showed no laser induced damage with fluences as high as 0.04 Jcm$^{-2}$ in 30fs at 800nm.  

\section{Characterization of Off-Axis Spiral Phase Mirrors} 
Several OASPM prototypes of various topological charge and diameter were manufactured using the University of Alberta's nanofabrication facility. Using electron beam evaporation, titanium and gold were deposited onto pre-polished glass mirror substrates of flatness $\lambda/10$. Titanium was used as an adhesion layer of roughly 25nm thick, followed by a 200nm layer of gold to ensure a high reflectivity across the entire surface. Sequential layers were then deposited on the mirror surface using a set of 5 aluminium masks. The masks were based on the designs of Sueda et al. \cite{Sueda:04} with modified sector angles according to Eq. (\ref{eq:12}). The modified angles were cut using a CNC water-jet from 3mm aluminium plate. The step deposition thickness was determined by Eq. (\ref{eq:14}). and monitored during deposition using a crystal thickness monitor. For a 16 step, $L=1$ OASPM designed to work at $45^\circ$ with a 632.8nm HeNe beam, we find the step sector angles given in Fig. 1d). and calculate each step height $h$ to be 27.97nm. 

Converting a Gaussian near-field beam to an OAM beam with integer topological charge $(L=\ell)$, will produce a far-field focal spot intensity given by the following formula \cite{Kotlyar:05,Longman:20},
\begin{equation} \label{eq6}
    I(r,\ell)=\frac{I_0\pi}{4}r^2e^{-r^2}\left|I_{\frac{\ell-1}{2}}\left(\frac{r^2}{2}\right)-I_{\frac{\ell+1}{2}}\left(\frac{r^2}{2}\right)\right|^2
\end{equation}
Here, $I_0$ is the peak intensity of the $\ell=0$ beam, $\ell$ is the azimuthal mode integer, $I_{n}(X)$ is the modified Bessel function of the first kind, $r$ is the normalized radius $r/w_0$ in the far-field plane where $w_0$ is the Gaussian beam waist parameter in the far-field defined as $w_0 = \lambda f/\pi R_0$. Here $\lambda$ is the wavelength of the laser, $f$ is the focal length of the lens, and $R_0$ is the Gaussian beam waist in the near-field.
Fig. 3a) and 3b) show the theoretical and experimental focal spots of a $R_0$ = 4.5mm, $\lambda$ = 632.8nm Gaussian beam focused using a $f$ = 750mm plano-convex lens. Fig. 3(c) and 3(d) shows the same beam reflected from a 2 inch diameter $45^\circ$, $N=16$, $L=1$ OASPM before focusing resulting in a high quality donut mode at focus. The topological charge of these beams was verifed using a Mach-Zehnder interferomtry \cite{Sueda:04}.

\begin{figure} \label{fig:3}
	\includegraphics{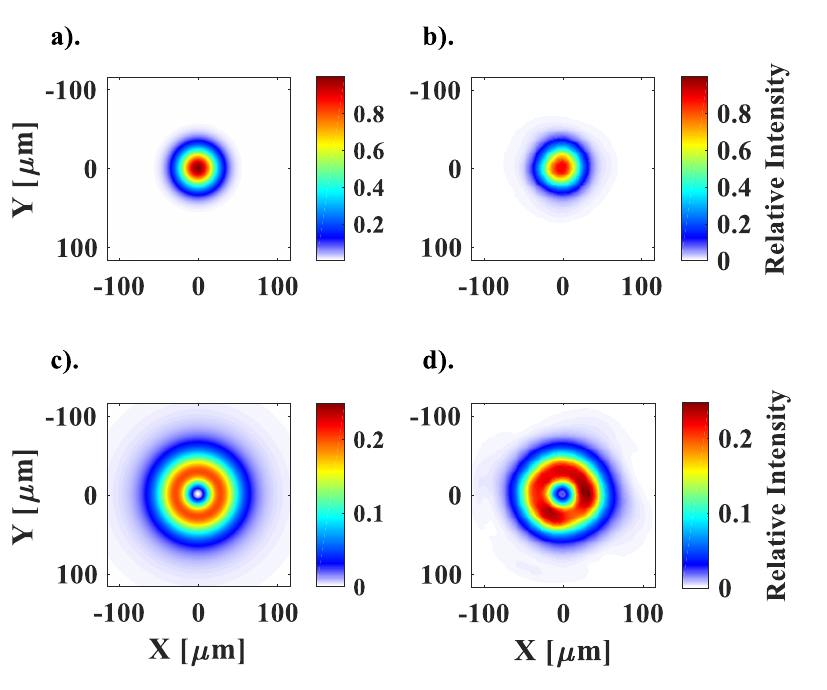}
	\caption{Theoretical and experimental focal spots of a Gaussian and an $L=1$ OAM beam from a collimated $R_0=4.5mm$ HeNe beam using a 750mm focal length lens. a) Theoretical Gaussian focal spot; b) Experimental Gaussian focal spot; c) Theoretical $\ell=1$ OAM focal spot; d) Experimental $L=1$ OAM focal spot.}
\end{figure}

Testing of the OASPM in a high power laser system was performed using the VEGA2 laser at the Center for Pulsed Lasers (CLPU) \cite{volpe}. The laser produces 3J super-Gaussian, 30fs pulses centred around 800nm with a near-field diameter of roughly 100mm. We can approximate the laser near-field to be roughly flat-top and as such can compare the laser focus to theoretical predictions. The generalized far-field intensity of a flat top beam carrying OAM can be written \cite{Longman:20}:
\begin{equation} \label{eq:18}
I(\rho,\ell)=\frac{I_{F}}{2}\left|\frac{\rho^\ell}{\ell!\left(\frac{\ell}{2}+1\right)}{}_{1}F_2\left(\frac{\ell}{2}+1;\ell+1,\frac{\ell}{2}+2;-\rho^2\right)\right|^2
\end{equation}Here, we introduce the peak intensity $I_F$ of the $\ell=0$ mode, the generalized hypergeometric function ${}_{1}F_2(a;b;z)$, and the normalized radius in the far-field plane $\rho = r/w_F$ where $w_F=\lambda f/\pi R_F$ in which $R_F$ is the radius of the flat top beam in the near-field. Various topological charge OASPM's were fabricated each with 16 steps $(N=16)$, to replace a 125mm diameter, $17.8^\circ$ incidence angle beamline mirror before an $f=400mm$, $40^\circ$ off axis parabolic focusing mirror. By fitting the near-field beam radius as $R_F$ = 30mm, we found good agreement between the predicted focal spot size and the measured. In this case, as in many high power laser systems, additional abberations in the beam prevent it from being perfectly diffraction limited. Fig. 4a), c), and e), show the theoretical focal spots for a laser beam with $\ell = 0, 1, 2$ respectively.  Fig. 4b), d), and f), show the measured focal spots with the corresponding peak intensities assuming the laser delivers 3J in 30fs to the focal spot. The peak intensity of the theoretical $\ell=0,1,2$ focal spots are calculated also assuming 3J in 30fs in the focal spot. From this it is clear to see a large discrepancy between the theoretical $\ell=0$ peak intensity and the experimental result, but with the higher order modes, we find better agreement between theory and experimental peak intensity. The images were taken by directly imaging the focal spot through a long working distance objective and with the laser operating in a low energy configuration (1mJ/shot), after which focal spot peak intensity was calculated assuming the total energy on focus.
\begin{figure} \label{fig:4}
	\includegraphics{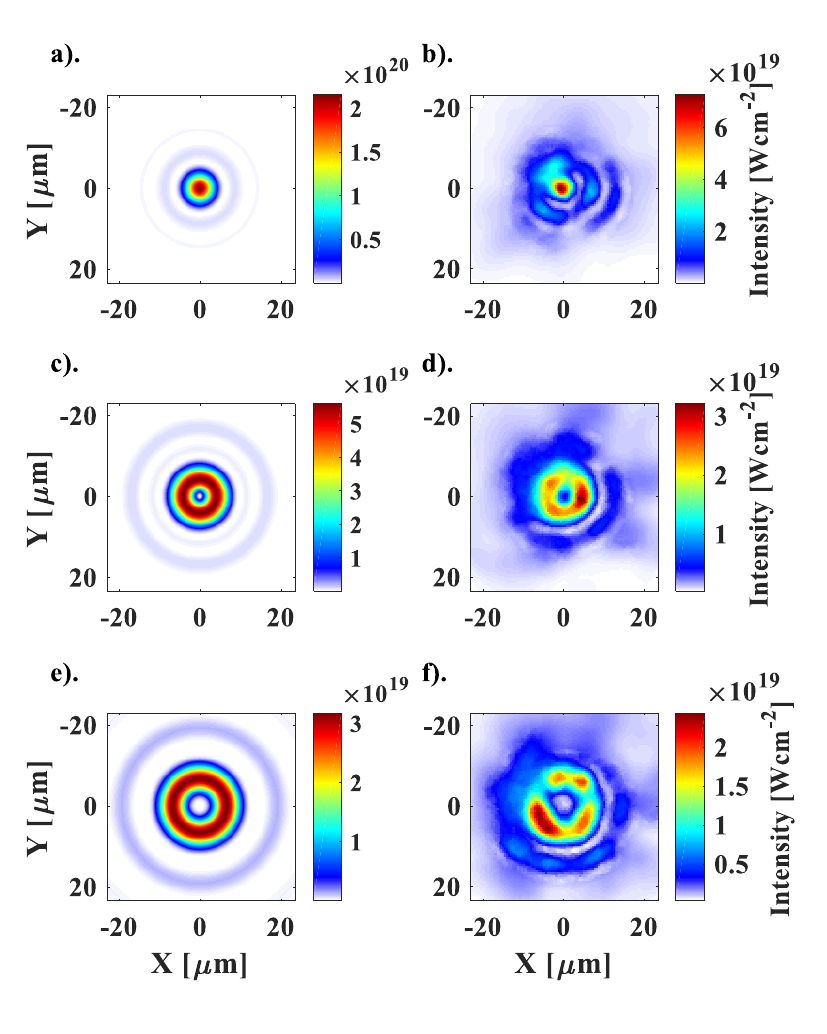}
	\caption{Theoretical and experimental focal spots generated with a 3J, 30fs laser. a). Theoretical focal spot $\ell = 0$. b). Experimental focal spot $L = 0$. c). Theoretical focal spot $\ell = 1$. d). Experimental focal spot $L = 1$. e). Theoretical focal spot $\ell = 2$. f). Experimental focal spot $L = 2$.}
\end{figure}
It is clear that there is an asymmetry in the experimental OAM focal spot rings. We believe this is primarily a consequence of spatial inhomogeneities in the laser intensity near-field profile, small contributions from defects in the OASPM, and residual focusing aberrations such as astigmatism. While we worked hard to minimize these effects on the beam, it was not possible to completely remove beam asymmetries in the given system.

\section{Conclusions}
In this paper, we have introduced and demonstrated that it is possible to efficiently mode convert a high-power laser into OAM beams using OASPM's. OASPM's are a cost effective, and simple method to mode convert almost any high-power laser to an OAM mode of desired topological charge. We have succesfully demonstrated the generation of OAM modes with charge $L=1,2$ using the OASPM with peak intensities above $2\times10^{19}Wcm^{-2}$. We found some asymmetries in our focused modes and believe this to be a result of near-field spatial inhomogenieties, manufacturing defects in the OASPM, and focusing aberrations such as astigmatism. We believe that these OAM modes are the highest intensity generated to date and that the OASPM is a tool that can open up the field of study of high-intensity OAM modes in the near future. 

\section*{Funding Information}
	This work was funded by the Natural Sciences and Engineering Research Council of Canada (NSERC) research grant number RGPIN-2014-05736, LaserLab Europe IV Grant No. 654148, and the Spanish Ministerio de Ciencia, Innovaci\'{o}n y Universidades (PALMA project FIS2016-81056-R) and Junta de Castilla y Le\'{o}n project CLP087U16.

\section*{Acknowledgments}
The authors thank H. Tiedje, K. McKee, H. Dexel, T. Kugler, R. Schwarze, A. Hryciw, and L. Schowalter for aiding in the manufacturing of the OASPM.  CLPU laser operators C. M\'{e}ndez, et al. and technical staff D. de Luis and D. Arana.

\bibliography{references}

\bibliographyfullrefs{references}

\end{document}